\documentclass[prl,twocolumn,superscriptaddress,showpacs,wasysym]{revtex4}
\def\beq{\begin{equation}}
\def\eeq{\end{equation}}
\usepackage{amsmath}
\usepackage{graphicx}
\usepackage{epstopdf}
\usepackage[usenames]{color}

\begin{document}

\title{Theromelectricity in Graphene: Effects of a gap and magnetic fields}

\author{Aavishkar A. Patel}
\affiliation{Department of Physics, Indian Institute of Technology, Kanpur, India}
\author{Subroto Mukerjee}
\affiliation{Department of Physics, Indian Institute of Science, Bangalore, India}

\begin{abstract}
We calculate the thermopower of monolayer graphene in various circumstances. First we show that experiments on the thermopower of graphene can be understood quantitatively with a very simple model of screening in the semiclassical limit. We can calculate the energy dependent scattering time for this model exactly. We then consider acoustic phonon scattering which might be the operative scattering mechanism in free standing films, and predict that the thermopower will be linear in any induced gap in the system. Further, the thermopower peaks at the same value of chemical potential (tunable by gate voltage) independent of the gap. Finally, we show that in the semiclassical approximation, the thermopower in a magnetic field saturates at high field to a value which can be calculated exactly and is independent of the details of the scattering. This effect might be observable experimentally.
\end{abstract}

\maketitle

Graphene has become one of the most studied electronic systems in recent years. The interest in this system stems from the existence of a relativistic band structure which gives rise to several interesting electrical properties~\cite{graphene_review}. Very high values of electron mobility combined with robust mechanical properties make graphene a promising candidate for use in several devices and technologies. Two recent developments have proven very significant in this regard: The ability to introduce a tunable gap in single sheets of graphene~\cite{balog,ajayan} and the fabrication of free standing sheets~\cite{shivaraman}. The former, in principle enables one to fabricate electronic devices using single sheets of graphene while the latter is an important step towards increasing the mobility of electrons by eliminating scattering effects from the substrate.

An enhanced mobility (and hence electrical conductivity) is important for thermoelectric applications as well. In addition to the technological significance of the phenomenon of thermoelectricity, the thermopower of a system is often used to shed light on various properties of the elementary charge carriers. These include the sign of their charge, the principle scattering mechanism and in certain limits, their entropy. Thus, the thermopower in graphene has also been studied extensively in the last few years both experimentally~\cite{kim,wei,ong} and theoretically~\cite{hwang1,dora,lofwander,yan,zhu,aji}. Interesting behavior such as the violations of the Mott formula at high temperatures have been found and attributed to the non-degenerate behavior of carriers. Most theoretical calculations have focussed on monolayer graphene without a gap with screened impurities being the dominant source of scattering. This is indeed the situation for graphene on substrates of SiO$_2$ on which most experiments have been carried out. However, with the two new developments mentioned in the previous paragraph, it has become important to consider the effect of a gap and other scattering mechanisms, which is one of the objectives of this paper.

A magnetic field can have interesting effects on thermoelectricity. It can produce oscillations~\cite{ong,zhu} or steps~\cite{mukerjee} in the thermopower under suitable circumstances which can provide information about the energy levels of the carriers of the system. The effect of magnetic fields on the thermopower of graphene has also been studied, mainly at low carrier densities, where the cyclotron frequency is high and oscillations have been seen indicative of the formation of Landau levels~\cite{kim,wei,ong,zhu}. However, not much attention has been paid to the effect of a magnetic field at higher values of filling when the cyclotron frequency is low and semiclassical physics can apply at reasonable temperatures. We show that in such a situation the thermopower can saturate at high field, an effect that might be observable experimentally.

In this paper, we consider semiclassical transport in graphene taking into account the effects of a gap, various scattering mechanisms and magnetic fields. We first show that it is possible to obtain quantitative agreement with thermopower measurements on graphene with a simple model of scattering in the semiclassical limit. While calculations along the same lines have been performed earlier~\cite{hwang1,dora,lofwander,yan,zhu,aji}, our calculation has the virtue of simplicity which allows us to calculate the scattering rate exactly and consequently make a fit to experimental data using only a weakly temperature dependent screening length and no other parameter. We then move on to consider the case of scattering due to acoustic phonons, which might be the dominant scattering mechanism in free standing films due to the absence of impurity effects from the substrate. Here, we find predict that the thermopower is proportional to the gap and obtain an exact expression in the limit of small gaps. This is quite distinct from what happens in the presence of substrates and might be used to characterize the gap. Finally, we calculate the effect of magnetic fields and find that in the semiclassical limit, the thermopower saturates to a value at high field, that can be calculated exactly, independent of the details of scattering. We show that for reasonable values of various parameters, this saturation can occur at a few Tesla of field, making it relevant experimentally.

Graphene has a hexagonal Brillouin zone. At the six corners of the zone, the dispersion is given by $E({\bf k}) = \pm \hbar v_f |{\bf k}|$
where $v_f$ is the fermi velocity.  The density of states is
\begin{equation}
g(E) = \frac{2|E|}{\pi\hbar^2v_{f}^2}
\label{Eq: DOS}
\end{equation}
In linear response response theory in the semiclassical approximation, the transport coefficients relevant to thermoelectric response at temperature $T$ are given by:
\begin{eqnarray} \nonumber
L^{11}_{ij} = \frac{e^2}{2\pi^2} \int \tau[E({\bf k})]\left[-\frac{\partial f}{\partial E({\bf k})}\right] v_i v_j  \, d{\bf k}, \\
L^{12}_{ij} = \frac{-e}{2\pi^2T} \int [E({\bf k})-\mu]\tau[E({\bf k})] \left[-\frac{\partial f}{\partial E({\bf k})} \right]v_i v_j  \, d{\bf k}.
\label{Eq:Lab}
\end{eqnarray}
where $v_i$ is the $i^{\rm th}$ component of the velocity, $\mu$, the chemical potential and $f$ the Fermi function. The thermopower
\begin{equation}
S = \frac{L^{12}_{xx}}{L^{11}_{xx}}.
\label{Eq:defS}
\end{equation}
The momentum integrals are carried out in the vicinity of all six corners of the Brillouin zone. The carrier density $\rho$ can be related to $\mu$ through the expression
\begin{equation}
\rho = \int_0^\infty g(E)f(E) \, dE - \int_{-\infty}^0 [1-f(E)]g(E) \, dE.
\label{Eq:murho1}
\end{equation}
$\rho$ is the excess carrier concentration over the completely filled lower band and $\mu=0$ when $\rho=0$.

As mentioned earlier, for graphene films on SiO$_2$ substrates the dominant scattering mechanism is due to charged impurities screened by the conduction electrons. While it is possible to obtain the screened dielectric function using techniques like the random phase approximation~\cite{hwang1,dora,lofwander,yan,zhu,aji}, we will assume a simple screened potential of the form
\begin{equation}
V(r) = \frac{qe}{4\pi\epsilon_{0}}\frac{e^{-r/\xi}}{r},
\label{Eq:screen}
\end{equation}
with a screening length $\xi$. As will be seen, a weakly temperature dependent $\xi$ enables us to make quantitative fits to the experimental data.  The main advantage of this potential is that the relaxation time can be calculated exactly using the Fermi golden rule yielding a fairly simple form
\begin{equation}
\frac{1}{\tau(k)} = \left(\frac{qe}{4\pi\epsilon_{0}}\right)^2\frac{\rho_{i}\xi^2k}{4\hbar^2v_{f}}\frac{1+2k^2\xi^2 - \sqrt{1+4k^2\xi^2}}{2k^4\xi^4},
\label{Eq:tau}
\end{equation}
where $q$ is the charge of the impurity, $\rho_{i}$ is the impurity concentration and $E(k)$ is given by Eqn.~\ref{Eq:disp}. When $\xi = \infty$, one obtains the known result $\tau(E) \propto |E|$~\cite{hwang2}.

\begin{figure}[h]
\includegraphics[width=3in]{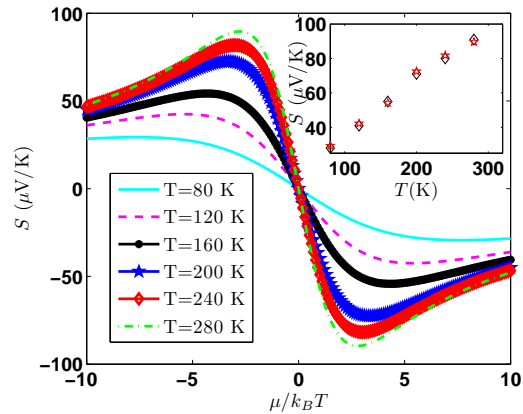}
\caption{Thermopower of graphene as a  function of chemical potential at different temperatures due to the scattering potential of Eqn.~\ref{Eq:screen}. The calculated peak values of the thermopower (diamonds) at every temperature can be directly compared to the experimental data (asterisks) of Ref.~\cite{ong} (inset). The only fitting parameter is the screening length $\xi$. The values $\xi=8.5$ nm for $T=80-160$ K and $\xi = 10$nm for $T=180-300$ K have been used here to obtain best fits to the experimental data.}
\label{Fig:screened}
\end{figure}

Eqn.~\ref{Eq:tau} can be used to calculate the thermopower. The required integrals have to be evaluated numerically and the thermopower as a function of $\mu$ at different temperatures is show in Fig.~\ref{Fig:screened}. The only parameter that is required to obtain this data is the screening length $\xi$, which we consider a fitting parameter. The thermopower has experimentally been measured as a function of gate voltage~\cite{ong}, which can be converted to $\mu$ only from knowledge of the sample dependent capacitance. However, the peak thermopower at every temperature can be read off directly and compared to experiments. We find very good agreement over a fairly large range of temperatures using a screening length of 8.5 nm for 80-160 K and 10 nm for 180-300 K.

The thermopower due to this screened potential can also be calculated in the presence of a gap $\Delta$ and does not show a very strong dependence on it at least for $\Delta \ll k_BT$. A much more interesting situation arises when the dominant scattering mechanism is acoustic phonon scattering. This mechanism is likely to be relevant for free-standing graphene films of the sort that have been fabricated recently. It can be shown from the Fermi golden rule that the relaxation time $\tau(E) \propto 1/|E|$ for this mechanism. Incidentally, this is also the energy dependence of $\tau(E)$ for localized impurity scattering. In the absence of a gap, the energy dependence of the product of the density of states with the square of the velocity in the integrals for $L^{\alpha \beta}_{xx}$ is exactly cancelled by that coming $\tau(E)$. Thus,
\begin{equation}
L^{12}_{xx} \propto \int_{-\infty}^{\infty} (E-\mu)\left(-\frac{\partial f}{\partial E}\right) \, dE = 0
\label{Eq:l_12_0}
\end{equation}
since the integrand is an odd function of $E - \mu$. The thermopower is identically zero with acoustic phonon scattering independent of the chemical potential (and hence carrier density)\footnote{Eventually the deviation of the dispersion from the form in Eqn.~\ref{Eq:disp} is going to give a nonzero thermopower. However, this value is going to be exponentially small in $W/k_BT$, where $W$ is the bandwidth.}. To demonstrate the effect of a gap, we assume  dispersion of the form
\begin{equation}
E({\bf k}) = \pm \sqrt{\hbar^2v_{f}^2|{\bf k}|^2 +\Delta^2}.
\label{Eq:disp}
\end{equation}
The magnitude of the velocity now picks up an energy dependence
\begin{equation}
v(E) = v_f \sqrt{1 - \frac{\Delta^2}{E^2}}.
\label{Eq:vel_energy}
\end{equation}
$g(E)$ is still given by Eqn.~\ref{Eq: DOS} and $\rho$ can be calculated in terms of $\mu$ to yield
\begin{eqnarray}\nonumber
\rho & = & \frac{2k_{B}^2T^2}{\pi\hbar^2v_{f}^2} \left\{{\rm Li}_{2}\left[-e^{-(\mu+\Delta)/k_{B}T}\right] - {\rm Li}_{2}\left[-e^{(\mu-\Delta)/k_{B}T}\right]\right\} \\
& & + \frac{2k_{B}T\Delta}{\pi\hbar^2v_{f}^2}\ln\frac{1+e^{(\mu-\Delta)/k_{B}T}}{1+e^{(-\mu-\Delta)/k_{B}T}},
\label{Eq:murho}
\end{eqnarray}
where ${\rm Li}_2(x)$ is the second polylogarithmic function.

The integrand in the expression for $L^{12}_{xx}$ is no longer an odd function of $E-\mu$. Further, the limits of integration also change. However, the form of $\tau(E)$ stays the same for acoustic phonon scattering due to $g(E)$ not changing. The resultant integration gives a non-zero value of the thermopower. This value can be obtained analytically in the limit of small gaps, $\Delta \ll \mu$ and $\Delta \ll k_{B}T$ and is
\begin{equation}
S = -\frac{k_{B}}{e}\frac{4\Delta}{k_BT}\frac{\mu}{k_BT} \frac{e^{-\mu/k_BT}}{(1+e^{-\mu/k_BT})^2}.
\label{Eq:thermo_gap}
\end{equation}

\begin{figure}[h]
\includegraphics[width=3in]{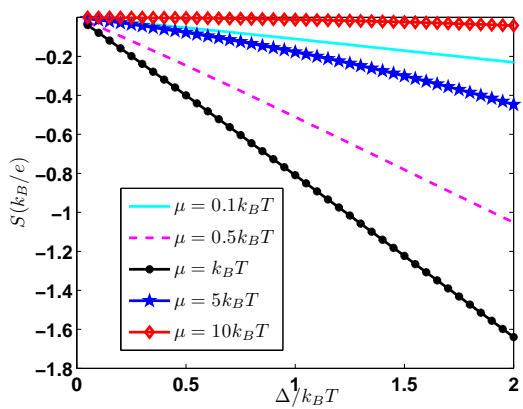}
\caption{Thermopower as a function of band gap for acoustic phonon scattering. The different curves correspond to different values of chemical potential.}
\label{Fig:acoustic1}
\end{figure}

\begin{figure}[h]
\includegraphics[width=3in]{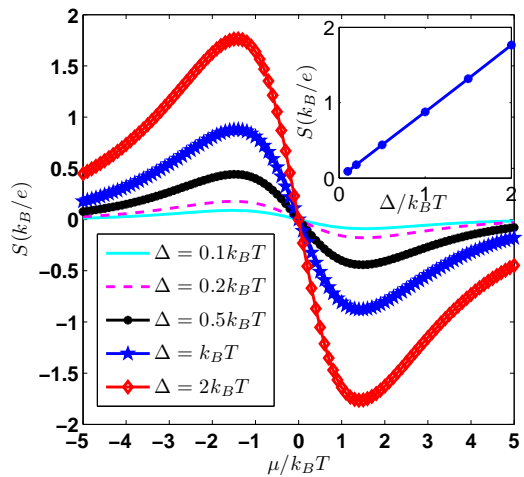}
\caption{Thermopower for different values of chemical potential for acoustic phonon scattering. The different curves correspond to different values of the gap. It can be seen that the peak occurs at approximately $\mu=1.5k_{B}T$ independent of the value of the gap. (Inset) The peak value of the thermopower as a function of the gap.}
\label{Fig:acoustic2}
\end{figure}

The thermopower can be calculated numerically for larger values of the gap. The results are shown in Fig.~\ref{Fig:acoustic1} from which it can be seen that the thermopower is roughly linear to reasonably large values of the gap. The thermopower as a function of chemical potential is shown in Fig.~\ref{Fig:acoustic2}. A curious fact is that the maximum appears at a value of $\mu/k_BT = 1.54$ roughly independent of the band gap. This can be seen directly for very low band gaps from Eqn.~\ref{Eq:thermo_gap} but holds more generally for larger gaps as well. The maximum value of the thermopower can be seen to be of the order of $k_B/e$ for sufficiently large values of the gap and is shown in the inset of Fig.~\ref{Fig:acoustic2}. It can also be seen that this peak thermopower is also inear in the gap up to far large values of the gap. Thus, the thermoelectric properties of a gapped free standing graphene sheet are likely to be quite different from those of a sheet on a substrate. In particular, the linear relation between the gap and the peak thermopower in the former case might provide a method to estimate the size of the gap.

We now focus our attention to the case of magnetothermal transport. The thermopower of graphene has been measured in the presence of a magnetic field at low temperature where oscillations have been observed as a function of the field consistent with quantization of energy levels~\cite{kim,ong}. Here we focus on semiclassical transport in the presence of a magnetic field which is relevant to experiments at higher temperature. The field $B$ is assumed to be perpendicular to the plane of the sheet. We consider the effect of magnetic fields in the absence of a gap. The solution of the semiclassical equations of motion gives orbits in which the energy of the electron is a constant. The linear response formulae Eqn.~\ref{Eq:Lab} then have to be modified to include the average of the velocities along these orbits. Operationally, this involves replacing one of the velocity operators by its average value in an orbit given by~\cite{ashcroft,wu2005}
\begin{equation}
\bar{\vec{v}}(\vec{k}) = \int_{-\infty}^{0} \frac{e^{t/\tau(E)}}{\tau(E)}\vec{v}(\vec{k}(t)) \, dt,
\label{Eq:avgv}
\end{equation}
\begin{figure}[h]
\includegraphics[width=3in]{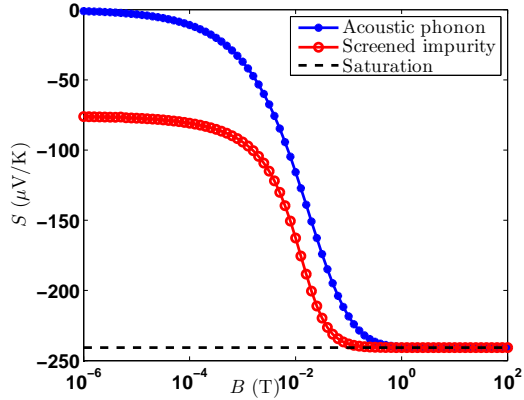}
\caption{Thermopower as a function of magnetic field for $T=273$ K and $\mu/k_BT = 1.59$, for acoustic phonon scattering and screened scattering (Eqn.~\ref{Eq:screen}). For acoustic phonon scattering, the deformation potential for graphene was taken to be 18 eV, the velocity of LA phonons $2.1\times10^6$ cm/s and mass density $7.6\times10^{-7}$ Kg/m$^2$. For screened impurity scattering, the impurities were assumed to have a charge of $2e$ and concentration of 1 ppm, and screening length of 10 nm.}
\label{Fig:field}
\end{figure}
where $\tau(E)$ is the relaxation time. It can be shown that the following relations result for the various transport coefficients when $\tau(E)$ explicitly depends on $|E|$:
\begin{eqnarray}\nonumber
L^{11}_{xx} & = & \frac{-e^2}{\pi\hbar^2}\int_{-\infty}^{\infty} E\frac{\partial f}{\partial E}\frac{dE}{\tau(E)\left[1/\tau(E)^2 + \beta^2/E^2\right]}, \\ \nonumber
L^{11}_{xy} & = & \frac{e^2\beta}{\pi\hbar^2}\int_{-\infty}^{\infty} \frac{\partial f}{\partial E}\frac{d|E|}{1/\tau(E)^2 + \beta^2/E^2}, \\ \nonumber
L^{12}_{xx} & = & \frac{e}{\pi\hbar^2T}\int_{-\infty}^{\infty} E(E-\mu)\frac{\partial f}{\partial E}\frac{dE}{\tau(E)\left[1/\tau(E)^2 + \beta^2/E^2\right]}, \\
L^{12}_{xy} & = & \frac{-e\beta}{\pi\hbar^2T}\int_{-\infty}^{\infty} (E-\mu)\frac{\partial f}{\partial E}\frac{d|E|}{1/\tau(E)^2 + \beta^2/E^2},
\label{Eq:Lab_mag}
\end{eqnarray}
where $\beta = ev_{f}^2B$\footnote{With a gap $\Delta$, all the integrands have to be multiplied by a factor of $\left( 1-\frac{\Delta^2}{E^2}\right)\Theta (E^2-\Delta^2)$.}. Note that while the integrals over the two bands for the diagonal coefficients $L^{\alpha \beta}_{xx}$ add, the ones for the off-diagonal coefficients $L^{\alpha \beta}_{xy}$ subtract. This is due to the fact that the sense of rotation of the orbits is opposite in the two bands. We have verified that the above integrals for screened impurity scattering yield values of conductivity and hall resistance as functions of $\mu$ that show very good qualitative agreement with the experimental data of Ref.~\cite{novoselov}. The Onsager relations $L^{\alpha \beta}_{xx} = L^{\alpha \beta}_{yy}$ and $L^{\alpha \beta}_{xy} = -L^{\alpha \beta}_{yx}$ can also be shown to hold. The thermopower in the presence of a magnetic field is given by the formula
\begin{equation}
S = \frac{L^{12}_{xx}L^{11}_{xx} + L^{12}_{xy}L^{11}_{xy}}{\left(L^{11}_{xx}\right)^2 + \left(L^{12}_{xy}\right)^2}.
\label{Eq:thermomag}
\end{equation}
\\ An analytic expression can be obtained for the saturation value of the thermopower at large magnetic fields.
\begin{widetext}
\begin{equation}
S_{xx} = \frac{1}{eT}\frac{4\mu k_{B}^2T^2{\rm Li}_{2}\left(-e^{\mu/k_{B}T}\right) - 12k_{B}^3T^3{\rm Li}_{3}\left(-e^{\mu/k_{B}T}\right) - 2\mu k_{B}^2T^2\pi^2/3}{4k_{B}^2T^2{\rm Li}_{2}\left( -e^{\mu/k_{B}T}\right) + \mu^2 + k_{B}^2T^2\pi^2/3},
\label{Eq:saturation}
\end{equation}
\end{widetext}
where Li$_3(x)$ is the third polylogarithmic function. For different scattering mechanisms, the field dependence of the thermopower has to be calculated numerically. This is shown for two scattering mechanisms, acoustic phonon scattering and screened impurity scattering in Fig.~\ref{Fig:field} for realistic values of the relevant parameters. The thermopower approaches the value given by Eqn.~\ref{Eq:saturation} for large enough fields in both cases. The saturation value is independent of the scattering mechanism as can be seen from Eqns.~\ref{Eq:Lab_mag}. For sufficiently large fields, $\beta/E$ dominates the factor $1/\tau(E)$ where the value of the integrand is appreciable. In this limit, $S \sim L^{12}_{xy}/L^{11}_{xy}$, which can be shown to be equal to Eqn.~\ref{Eq:saturation}. Effects of the Zeeman splitting of the electrons can be neglected even up to fields of 10 T since the corresponding energy is still much smaller compared to the Fermi energy. For the parameters used, the saturation field is of the order of a few Tesla as can be seen from Fig.~\ref{Fig:field}, which could make the effect experimentally observable.

We thank Arindam Ghosh for helpful discussions. We would like to acknowledge financial support from the Department of Science and Technology (DST) of the government of India especially through the Kishore Vigyan Protsahan Yojna (AAP) and the Ramanujan Fellowship (SM).

\bibliographystyle{apsrev1}
\bibliography{graphene}

\end{document}